\documentclass[aps,prd,twocolumn,showkeys,amsmath,amssymb,superscriptaddress]{revtex4}
\usepackage{graphicx}
\usepackage{epstopdf}   
\usepackage{multirow}
\usepackage{subfigure}
\usepackage{extarrows}
\usepackage{feynmf}
\usepackage{enumitem}
\usepackage[colorlinks,citecolor=blue,anchorcolor=red,menucolor=red,linkcolor=red,filecolor=red,runcolor=red,urlcolor=blue,frenchlinks=red]{hyperref}
\usepackage{color}

\newcommand{\feynp}[1]{#1\kern-0.45em/}

\def\FF(s){\left[(\alpha+\beta)m_c^2-\alpha\beta s\right]}
\def\HH(s){\left[m_c^2-\alpha(1-\alpha) s\right]}
\def\KK(s){\left[\gamma m_c^2-\gamma(1-\gamma) s\right]}

\allowdisplaybreaks[3]

\begin{document}

\title{The decay properties of two- and three- gluon glueballs}

\author{Wei-Han Tan}
\affiliation{School of Physics, Southeast University, Nanjing 211189, China}

\author{Hua-Xing Chen}
\email{hxchen@seu.edu.cn}
\affiliation{School of Physics, Southeast University, Nanjing 211189, China}

\author{Ding-Kun Lian}
\email{liandk@seu.edu.cn}
\affiliation{School of Physics, Southeast University, Nanjing 211189, China}

\author{Wen-Ying Liu}
\email{liuwenying@lzu.edu.cn}
\affiliation{School of Physical Science and Technology, Lanzhou University, Lanzhou 730000, China}
\affiliation{Lanzhou Center for Theoretical Physics, Key Laboratory of Theoretical Physics of Gansu Province, Key Laboratory of Quantum Theory and Applications of MoE, Gansu Provincial Research Center for Basic Disciplines of Quantum Physics, Lanzhou University, Lanzhou 730000, China}
\affiliation{MoE Frontiers Science Center for Rare Isotopes, Lanzhou University, Lanzhou 730000, China}
\affiliation{Research Center for Hadron and CSR Physics, Lanzhou University and Institute of Modern Physics of CAS, Lanzhou 730000, China}

\begin{abstract}

We previously studied the decay properties of two- and three-gluon glueballs using the Fierz rearrangement method and obtained their relative branching ratios in Ref.~\cite{Tan:2026uue}. In this work, we extend and develop this analysis by providing a more complete treatment of two-gluon glueball decays and by considering the tensor and pseudotensor states with \(J^{PC}=2^{++}\) and \(2^{-+}\). We also perform an independent QCD sum rule analysis of the \(0^{++}\) two-gluon glueball  decay. The consistency between the two approaches provides a useful check of the Fierz analysis. Our results support a sizable gluon component in the \(f_0(1710)\) and favor the \(0^{-+}\) glueball interpretation of the \(\eta(2370)\). For the tensor glueball, the vector--vector ($VV$) decay channels, especially \(K^{*}(892)\bar{K}^{*}(892)\), are found to be favorable for experimental searches. We also study three-gluon glueballs with \(J^{PC}=0^{++}\) and \(1^{+-}\) and identify several potentially favorable three-meson decay channels, including \(\pi\pi\omega\) and \(K\bar K\phi\). These results provide possible guidance for future experimental searches for glueball states.

\end{abstract}

\maketitle

\section{Introduction}
\label{sec:intro}
Glueballs, color-singlet bound states composed entirely of gluons through their self-interactions, represent one of the most distinctive predictions of Quantum Chromodynamics (QCD) and provide an important window into its nonperturbative dynamics~\cite{pdg}. Lattice QCD and a variety of phenomenological approaches consistently predict the lightest scalar glueball to have a mass around \(1.6~\mathrm{GeV}\), while the lightest pseudoscalar and tensor glueballs are expected to lie at higher masses~\cite{Carlson:1982er,DeGrand:1975cf,Isgur:1983wj,Isgur:1984bm,Szczepaniak:1995cw,Llanes-Estrada:2005bii, Mathieu:2008bf,Chen:2005mg,Forkel:2007ru,Li:2013oda,Brunner:2015yha,Michael:1988jr,Yamanaka:2019yek,Richards:2010ck,Ye:2012gu,Morningstar:1999rf,Meyer:2004gx,Gregory:2012hu,Athenodorou:2020ani,Sarantsev:2021ein,Karch:2006pv,Chen:2021bck}. Over the past decades, a number of experimentally observed resonances, including the \(f_0(1500)\), \(f_0(1710)\), \(f_2(2340)\), \(\eta(2370)\), and \(X(2600)\), have been proposed as possible glueball candidates~\cite{BaBar:2018uqa,BaBar:2021fkz,LHCb:2025nys,BESIII:2022sfx,BESIII:2023wfi,BESIII:2026mvn}. Nevertheless, the identification of glueballs remains challenging because of their possible mixing with conventional quark--antiquark mesons and the model dependence inherent in many theoretical descriptions. We refer the reader to Refs.~\cite{Klempt:2007cp,Crede:2008vw,Mathieu:2008me,Meyer:2010ku,Meyer:2015eta,Ochs:2013gi,Brambilla:2014jmp,Sonnenschein:2016pim,Briceno:2017max,Guo:2017jvc,Bass:2018xmz,Ketzer:2019wmd,Roberts:2021nhw,Fang:2021wes,Jin:2021vct,Gross:2022hyw,Chen:2022asf,Amsler:1995tu,Qin:2017qes,Klempt:2021wpg,Wang:2026iyq,Cotanch:2005ja,Lodha:2024qby,Lodha:2024yfn} for comprehensive reviews.

In our previous work, we proposed a framework based on the Fierz rearrangement~\cite{Fierz:1937wjm} to investigate glueball decays through the algebraic structures dictated by QCD symmetries. Within this approach, glueball decays are described by converting constituent gluons into quark--antiquark currents, followed by exact Fierz rearrangements in both color and Lorentz spaces. In this work, we continue to employ the same framework under the simplifying assumptions like the spirit of the constituent quark model~\cite{Gell-Mann:1964ewy,Zweig:1964ruk,Zweig:1964jf,Isgur:1978xj,DeRujula:1975qlm,Godfrey:1985xj} that the decay is dominated by constituent gluons, each gluon excites a quark--antiquark pair, mesons are described as quark--antiquark bound states, and final-state interactions are neglected. Although this treatment relies on an effective constituent picture rather than a complete dynamical description, the Fierz rearrangement itself follows directly from the color and Lorentz structures of QCD. It therefore allows us to examine the flavor and Lorentz structures of glueball decays without introducing a detailed model for the decay dynamics. We regard this as a useful feature of the approach, while keeping in mind that dynamical effects beyond the constituent-level treatment may modify the detailed decay patterns.

In this work, we develop a unified Fierz-rearrangement analysis of two- and three-gluon glueball decays. For the two-gluon systems, we further develop the previous analysis by providing more complete derivations and numerical details for the \(J^{PC}=0^{++}\) and \(0^{-+}\) states, and extend the study to the tensor and pseudotensor states with \(J^{PC}=2^{++}\) and \(2^{-+}\). For the tensor and pseudotensor states, we employ the interpolating currents proposed in Ref.~\cite{Chen:2021bck} and evaluate their two-body decay patterns and relative branching ratios. The resulting vector--vector \((VV)\) final states decay channels may provide useful targets for future experimental searches. We also provide a more detailed treatment of the three-gluon glueballs with \(J^{PC}=0^{++}\) and \(1^{+-}\). We illustrate the decay mechanism through two successive Fierz rearrangements leading to three-meson final states, and give a more complete presentation of the corresponding analysis and results obtained previously.

In addition to the Fierz rearrangement analysis, we investigate the decay of the scalar two-gluon glueball within the framework of QCD sum rules by employing three-point correlation functions. Since the two approaches describe the decay process from different perspectives, a comparison between them offers an opportunity to examine the consistency of the corresponding predictions. Although the QCD sum rule analysis presented here is limited to the scalar channel, the overall agreement obtained for the dominant decay modes suggests that the Fierz rearrangement approach is able to capture the main features of glueball decays at the present level of analysis and may provide a useful guide for studying the other two- and three- gluon glueballs.

The remainder of this paper is organized as follows. In Sec.~\ref{sec:current}, we introduce the interpolating currents adopted for the glueball states considered in this work. We then develop the Fierz rearrangement formalism and derive the corresponding decay amplitudes in Sec.~\ref{sec:Fierz}. Based on this framework, we discuss the decay properties of scalar, pseudoscalar, tensor, and pseudotensor two-gluon glueballs in Sec.~\ref{sec:Result1}, followed by an analysis of three-gluon glueball decays in Sec.~\ref{sec:Result2}.  In Sec.~\ref{sec:Sum rule}, we subsequently present the QCD sum rule study of the scalar two-gluon glueball and compare its predictions with those obtained from the Fierz rearrangement method. Finally, we summarize our main results and comment on possible directions for future investigations in Sec.~\ref{sec:Summary}.

\section{Glueball interpolating currents}
\label{sec:current}

The interpolating currents of two-gluon and three-gluon glueballs have been systematically investigated in Ref.~\cite{Chen:2021bck}. In the present work, we focus on the two-gluon glueballs with quantum numbers $J^{PC}=0^{++}/0^{-+}/2^{++}/2^{-+}$ and three-gluon glueballs with quantum numbers $J^{PC}=0^{++}/1^{+-}$, which are denoted by $|\mathrm{GG};\,0^{++}/0^{-+}/2^{++}/2^{-+}\rangle$ and $|\mathrm{GGG};\,0^{++}/1^{+-}\rangle$, respectively. Their corresponding interpolating currents are chosen as
\begin{align}
J_0 &= g_s^2 G^{\mu \nu}_i G_{\mu \nu}^i  \, , 
\\
\label{eq:J0}
\widetilde{J}_0 &= g_s^2 G^{\mu \nu}_i \widetilde{G}_{\mu \nu}^i  \, , 
\\
J_2^{\alpha_1 \alpha_2,\beta_1 \beta_2} &= \mathcal{S}[ g_s^2 G^{\alpha_1 \beta1}_i G^{i,\alpha_2 \beta2} ]\, , 
\\
\widetilde{J}_2^{\alpha_1 \alpha_2,\beta_1 \beta_2} &= \mathcal{S}[ g_s^2 G^{\alpha_1 \beta1}_i \widetilde{G}^{i,\alpha_2 \beta2} ]\, ,
\\
\eta_0 &= f^{ijk}g_s^3 G^{\mu \nu}_i G_{j,\nu\rho}  G_{k,\mu}^{\rho}\, , 
\\
\eta_1^{\alpha\beta} &= d^{ijk}g_s^3 G^{\mu \nu}_i G_{j,\mu\nu}  G_{k}^{\alpha\beta}\, ,
\end{align}
where $i,j,k$ denotes the color indices (running from 1 to 8), and $\mu,\nu, \rho,\alpha_i,\beta_i$ are Lorentz indices; $f^{ijk}$ and $d^{ijk}$ are the antisymmetric and symmetric structure constants, respectively; $\mathcal{S}[...]$ denotes symmetrization and subtracting trace terms in the two sets ${\alpha_1,\alpha_2}$ and ${\beta_1,\beta_2}$ simultaneously. For practical calculations, we can implement the symmetrization operation $\mathcal{S}$ through the product of two symmetric-traceless projectors acting independently on the $\alpha$- and $\beta$-index spaces. The resulting projection operator reads 

\begin{align}
& \Gamma^{\alpha_1 \beta_1 \alpha_2 \beta_2, \rho_1 \sigma_1 \rho_2 \sigma_2}
 = \label{eq:project}\\ 
 \nonumber & \Bigl[
\frac{1}{2}\bigl(g^{\alpha_1\rho_1}g^{\alpha_2\rho_2}
+g^{\alpha_1\rho_2}g^{\alpha_2\rho_1}\bigr)
-\frac{1}{4}\,g^{\alpha_1\alpha_2}g^{\rho_1\rho_2}
\Bigr] \times \\ 
\nonumber & \Bigl[
\frac{1}{2}\bigl(g^{\beta_1\sigma_1}g^{\beta_2\sigma_2}
+g^{\beta_1\sigma_2}g^{\beta_2\sigma_1}\bigr)
-\frac{1}{4}\,g^{\beta_1\beta_2}g^{\sigma_1\sigma_2}
\Bigr] \, .
\end{align}

The gluon field-strength tensor $G_{\mu\nu}^{i}$ can be expressed in terms of the gauge field $A_{\mu}^{i}$ as
\begin{equation}
G_{\mu\nu}^{i} = \partial_{\mu} A_{\nu}^{i} - \partial_{\nu} A_{\mu}^{i} + g_sf^{ijk} A_{\mu}^{j}A_{\nu}^{k} \, .
\end{equation}
together with its dual tensor $\widetilde{G}_{\mu\nu}^i =G^{i,\rho\sigma} \times \epsilon_{\mu\nu\rho\sigma}/2$.

These six currents can be coupled, respectively, to the six glueball states $|\mathrm{GG};\,0^{++}/0^{-+}/2^{++}/2^{-+}\rangle$ and $|\mathrm{GGG};\,0^{++}/1^{+-}\rangle$. The couplings of these currents to the corresponding glueball states are expressed by:
\begin{align}
\langle 0 | J_0 | \mathrm{GG};\,0^{++} \rangle &= f_{0^{++}} \, , 
\\
\langle 0 | \widetilde{J}_0 | \mathrm{GG};\,0^{-+} \rangle &= f_{0^{-+}}  \, , 
\\
\langle 0 | J_2^{\cdots} | \mathrm{GG};\,2^{++} \rangle &= if_{2^{++}} \mathcal{S}[\epsilon^{\alpha_i\beta_i\mu_i\nu_i}p_{\nu_i}]^2 \epsilon_{\mu_1 \mu_2} \, , 
\\
\langle 0 | \widetilde{J}_2^{\cdots} | \mathrm{GG};\,2^{-+} \rangle &= if_{2^{-+}} \mathcal{S}[\epsilon^{\alpha_i\beta_i\mu_i\nu_i}p_{\nu_i}]^2 \epsilon_{\mu_1 \mu_2} \, ,
\\
\langle 0 | \eta_0 | \mathrm{GGG};\,0^{++} \rangle &= f^{'}_{0^{++}} \, ,
\\
\langle 0 | \eta^{\alpha\beta}_1 | \mathrm{GGG};\,1^{+-} \rangle &= f_{1^{+-}} \epsilon^{\alpha\beta\mu\nu} \epsilon_{\mu} p_{\nu} \, ,
\end{align}
where $f$ denotes the decay constants of these glueball states, $\epsilon^{\alpha\beta\mu\nu}$ is the four-dimensional Levi-Civita tensor with the convention \(\epsilon^{0123}=+1\), \(\epsilon_{\mu}\) and \(\epsilon_{\mu\nu}\) represent the polarization vector and the symmetric traceless polarization tensor, respectively, and \(\mathcal{S}[\cdots]\) denotes symmetrization together with the subtraction of trace terms in the two index sets ${\alpha_i,\alpha_2}$ and ${\beta_1,\beta_2}$ simultaneously with
\begin{align}
[\cdots]^N &= \epsilon^{\alpha_1\beta_1\mu_1\nu_1}p_{\nu_1} \cdots \epsilon^{\alpha_N\beta_N\mu_N\nu_N}p_{\nu_N} \, . 
\end{align}

Following the constituent-gluon picture adopted in our previous analyses, the gluon degrees of freedom are assumed to couple to vector quark-antiquark currents through
\begin{equation}
A_{\mu}^{i} \rightarrow \lambda^{i}_{ab}  \times \bar{q}^{a} \gamma_{\mu} q^b\, ,
\label{eq:Amu}
\end{equation}
where $\lambda^{i}_{ab}$ are the Gell-Mann matrices and $a,b$ denote fundamental color indices. This prescription allows the gluon operators to be rewritten in terms of quark bilinears, after which color and Dirac Fierz rearrangements can be performed to identify the relevant meson-meson decay channels.

\section{Overview of the Fierz rearrangement formalism}
\label{sec:Fierz}

\begin{figure*}[]
\begin{center}
\subfigure[]{
\scalebox{0.4}{\includegraphics{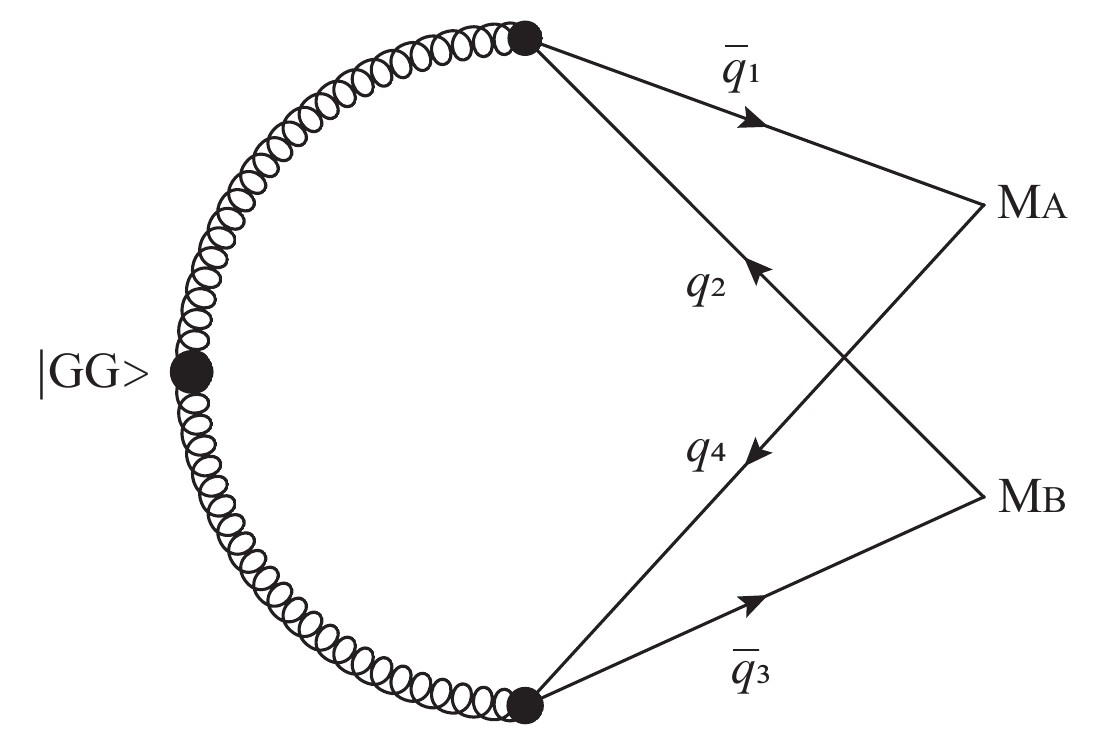}}}~~~~~
\subfigure[]{
\scalebox{0.4}{\includegraphics{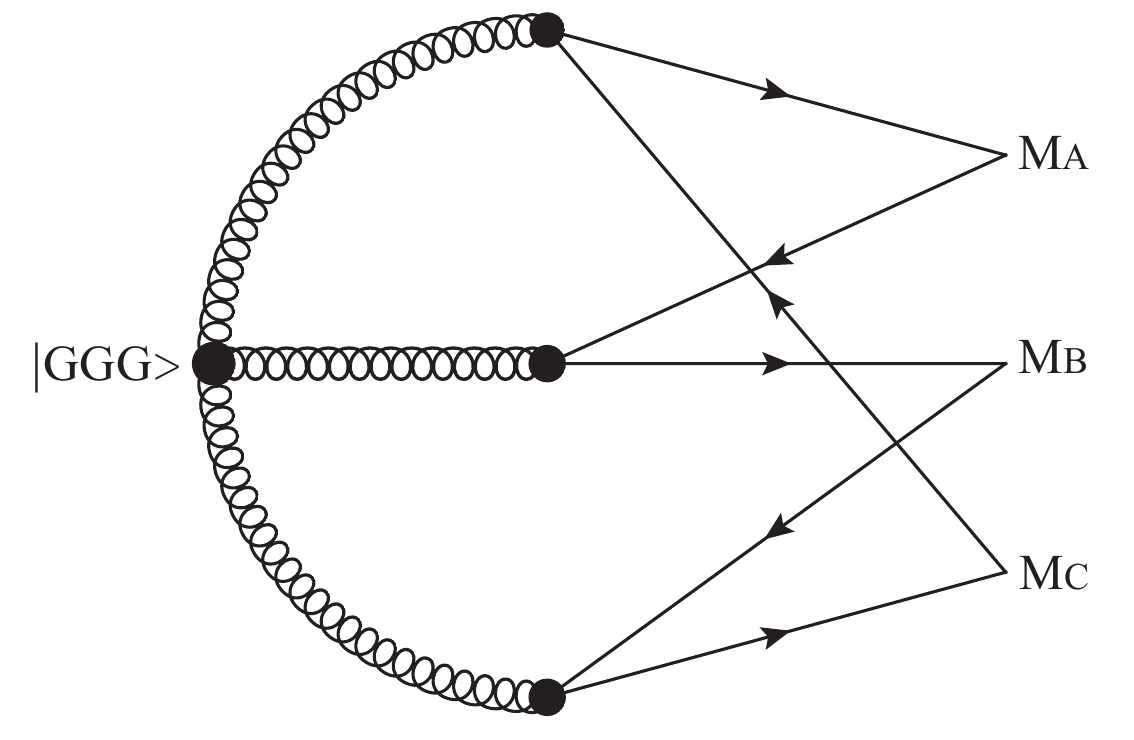}}}~~~~~
\end{center}
\caption{Diagram (a) illustrates the schematic diagram of a two-gluon glueball decaying into two mesons via a single Fierz rearrangement, while Diagram (b) depicts the schematic diagram of a three-gluon glueball decaying into three mesons through two successive Fierz rearrangements.}
\label{fig:Fierz}
\end{figure*}

In this section, we outline the Fierz rearrangement formalism adopted throughout the present work. We apply the same framework to both two-gluon and three-gluon glueballs, although the explicit decompositions depend on the quantum numbers and constituent structures of the corresponding interpolating currents. For two-gluon glueballs, one Fierz rearrangement in both color and Dirac spaces is sufficient to reorganize the operators into color-singlet hadronic configurations. The three-gluon currents follow the same general strategy but require an additional rearrangement because of the extra constituent gluon. Since the underlying procedure is common to all cases, we illustrate it below using the scalar current \(J_0\), which provides the simplest example. Fig.~\ref{fig:Fierz} summarizes the corresponding rearrangement of two- and three-gluon glueballs.

We begin with the interpolating current and first rewrite the gluon fields in terms of quark bilinears through the effective gluon--quark coupling introduced in Eq.~(\ref{eq:Amu}). We then perform the color and Dirac-space Fierz rearrangements successively, which reorganize the resulting four-quark operators into products of color-singlet meson currents. The procedure may be summarized schematically as
\begin{align}
|\mathrm{GG};0^{++}\rangle & \xleftrightarrow{~~~~~} J_0 = G^{\mu \nu}_i \times G_{\mu \nu}^i \times g_s^2
\label{eq:expand1} \\
\nonumber &\xrightarrow{~~~~~} ({M_{A}})_{\alpha}^{i} \times ({M_{B}})_{\beta}^{i} \times \mathbb{C}_1 + \cdots \\
\nonumber &\xrightarrow{~~~~~} \lambda^{i}_{ab} \lambda^{i}_{cd} \times \bar{q}_1^{a} \gamma_{\alpha} q_2^b \times \bar{q}_3^{c} \gamma_{\beta} q_4^d \times \mathbb{C}_2 + \cdots \\
\nonumber &\xrightarrow{\rm color} \delta_{ad} \delta_{cb} \times \bar{q}_1^{a} \gamma_{\alpha} q_2^b \times \bar{q}_3^{c} \gamma_{\beta} q_4^d \times \mathbb{C}_3 + \cdots \\
\nonumber &\xrightarrow{\rm Fierz} \delta_{ad} \delta_{cb} \times \bar{q}_1^{a} \Gamma_{1} q_4^d \times \bar{q}_3^{c} \Gamma_{2} q_2^b \times \mathbb{C}_4 + \cdots \\
\nonumber &\xrightarrow{~~~~~} [\bar{q}_1^{a} \Gamma_{1} q_4^a]_A \times [\bar{q}_3^b \Gamma_{2} q_2^b]_B \times \mathbb{C}_4 + \cdots \, .
\end{align}
Eq.~(\ref{eq:expand1}) highlights the sequence of transformations that connects the glueball interpolating current with color-singlet meson operators. At each stage, the coefficients $\mathbb{C}_{1,2,3,4}$ collect the corresponding numerical factors, while the matrices $\Gamma_{1,2}$ denote the Dirac structures generated by the Lorentz-space Fierz rearrangement. Throughout the discussion, we take quark fields $\bar{q}_1^{a}$ and $q_2^{b}$ to have the same flavor, and likewise for the quark fields $\bar{q}_3^{c}$ and $q_4^{d}$. The subscripts $M_A,M_B$, labeling the two meson components, are introduced for convenience in the discussion below. To carry out the color rearrangement, we employ the SU(3) identity
\begin{equation}
\lambda^{i}_{ab}\lambda^{i}_{cd} = 2\delta_{ad}\delta_{cb}-\frac{2}{3}\delta_{ab}\delta_{cd}\, ,
\end{equation}
and we follow the standard Lorentz-space Fierz identities collected in Ref.~\cite{Chen:2016qju}. Once these rearrangements have been completed, we replace derivatives acting on quark bilinears by the momenta of the corresponding mesons according to
\begin{equation}
\partial^\mu (\bar q \Gamma q)_A
\rightarrow
p_A^\mu (\bar q \Gamma q)_A.
\end{equation}
which allows us to express the decay amplitudes in terms of hadron matrix elements.

The complete decomposition of the scalar current contains nineteen independent operator structures. Rather than interrupting the discussion with a lengthy expression, we collect the full result in Appendix~\ref{app:Fierz} and display only several representative terms here to illustrate the structures generated by the Fierz rearrangement:
\begin{eqnarray}
J_0 &=& +\frac{1}{4}[\bar{u}_1^a \gamma_{\mu}\gamma_5 d_4^a]_A [\bar{d}_3^b \gamma_{\mu}\gamma_5 u_2^b]_B \times(q_A^{\nu}+q_B^{\nu})^2 
\label{eq:fierz1}
\\ \nonumber & & +\frac{1}{2}[\bar{u}_1^a \gamma_{\mu}\gamma_5 d_4^a]_A [\bar{d}_3^b \gamma_{\nu}\gamma_5 u_2^b]_B \times (q_A^{\mu}+q_B^{\mu})(q_A^{\nu}+q_B^{\nu})
\\ \nonumber & & +\frac{3}{4}[\bar{u}_1^a \gamma_5 d_4^a]_A [\bar{d}_3^b \gamma_5 u_2^b]_B\times (q_A^2+q_B^2+2q_A \cdot  q_B)
\\ \nonumber & & +\frac{1}{4}[\bar{u}_1^a \gamma_{\mu} d_4^a]_A [\bar{d}_3^b \gamma_{\mu} u_2^b]_B \times(q_A^{\nu}+q_B^{\nu})^2
\\ \nonumber & & +\frac{1}{2}[\bar{u}_1^a \gamma_{\mu} d_4^a]_A [\bar{d}_3^b \gamma_{\nu} u_2^b]_B \times (q_A^{\mu}+q_B^{\mu})(q_A^{\nu}+q_B^{\nu})
\\ \nonumber & & -\frac{1}{2}[\bar{u}_1^a \sigma_{\mu\alpha} d_4^a]_A [\bar{d}_3^b \sigma_{\nu\alpha} u_2^b]_B \times (q_A^{\mu}+q_B^{\mu})(q_a^{\nu}+q_b^{\nu})
\\ \nonumber & & +\frac{1}{8}[\bar{u}_1^a \sigma_{\mu\nu} d_4^a]_A [\bar{d}_3^b \sigma_{\mu\nu} u_2^b]_B \times (q_A^2+q_B^2+2q_A \cdot  q_B)\, .
\end{eqnarray}

We follow the same strategy for the remaining interpolating currents considered in this work. For the tensor and pseudotensor two-gluon currents, additional symmetric-traceless projections appear because of their Lorentz structures, making the intermediate expressions considerably more involved. The three-gluon currents can also be analyzed within the same framework. In that case, we carry out a second Fierz rearrangement before reorganizing the operators into color-singlet hadronic configurations. Since our primary goal is to discuss the resulting decay properties, rather than present lengthy intermediate algebra, we leave the detailed decomposition of the three-gluon currents to the corresponding phenomenological section and summarize only the expressions required for the subsequent analyses.

Once the Fierz rearrangement has been completed, we identify the resulting color-singlet quark bilinears with physical meson states through the standard current--meson matrix elements. Throughout this work, we employ the ground-state meson couplings summarized in Table~\ref{tab:coupling}. For completeness, we note that the corresponding relations for the excited meson operators have already been derived in Ref.~\cite{Tan:2025nir}; to avoid unnecessary repetition, we do not reproduce them here. Combining these current--meson matrix elements with the Fierz decomposition allows us to rewrite the quark-level operators in terms of hadronic decay amplitudes. In the present work, we therefore focus on extending the phenomenological analysis and applying the same framework to the additional glueball systems considered in this paper.

Before proceeding to the numerical analyses, we briefly comment on the scope of the present framework. The Fierz rearrangement determines the Lorentz, flavor, and color structures of the decay amplitudes by matching the interpolating currents onto meson operators within a constituent-gluon picture. In this sense, the resulting amplitudes should be regarded as effective hadronic couplings determined primarily by their underlying algebraic structures. At this analysis, we do not include dynamical effects associated with the spatial wave-function overlaps of the hadrons~\cite{Moshinsky:1959qbh,Ribeiro:1981fk,Thompson:1988zz}, nor do we consider possible final-state interactions after the hadronization process. Consequently, the overall normalization of the decay amplitudes, together with channel-dependent dynamical corrections, remains beyond the scope of the present treatment. Instead, we concentrate on relative decay widths and branching ratios, for which many of these common uncertainties are expected to be significantly reduced. Within these approximations, the Fierz rearrangement provides a convenient framework for comparing the decay patterns of different glueball candidates and for confronting the resulting predictions with other theoretical approaches.

\begin{table*}[hbt]
\begin{center}
\renewcommand{\arraystretch}{1.36}
\caption{Couplings of meson operators with quark contents $\bar q q/\bar q s/\bar s s (q=u/d)$ to meson states, where color indices are omitted for simplicity. All the isovector meson operators have the quark content $\bar q \Gamma q = \left(\bar u \Gamma u - \bar d \Gamma d\right)/\sqrt2$, and all the isoscalar light meson operators have the quark content $\bar q \Gamma q = \left(\bar u \Gamma u + \bar d \Gamma d\right)/\sqrt2$. Here, \(\sigma\) denotes \(f_0(500)\), \(\kappa\) denotes \(K_0^*(700)\), \(f_0\) denotes \(f_0(980)\), and \(K^*\) denotes \(K^*(892)\).}
\begin{tabular}{ c | c | c | c | c | c}
\hline\hline
~Operators--~ & ~~$(I^G)J^{PC}$~~ & ~~~~~~~~Mesons~~~~~~~~ & ~~$(I^G)J^{PC}$~~ & ~~~Couplings~~~ & ~~~~~~~~Decay Constants~~~~~~~~
\\ \hline\hline
$\bar q q$                & $(0^+)0^{++}$                  & $\sigma$ & $(0^+)0^{++}$ & $\langle 0 | J | \sigma \rangle = m_\sigma f_\sigma$          & $f_\sigma \sim 380$~$\mathrm{MeV}$~
\\ \hline
\multirow{2}{*}{$\bar q i\gamma_5 q$ }     &  \multirow{2}{*}{$(0^+)0^{-+}$}    & $\eta$    & $(0^+)0^{-+}$ & $\langle 0 | J | \eta \rangle = \lambda_{\eta_q}$        & $\lambda_{\eta_q} = 172466$~$\mathrm{MeV}^2$~
\\ \cline{3-6} &      & $\eta^\prime$          & $(0^+)0^{-+}$ & $\langle 0 | J | \eta^\prime \rangle = \lambda_{\eta^\prime_q}$
 &  $\lambda_{\eta^\prime_q} = 119280$~$\mathrm{MeV}^2$~
\\ \hline
$\bar q \gamma_\mu q$ & $(0^-)1^{--}$                  & $\omega$             & $(0^-)1^{--}$ & $\langle 0 | J_\mu | \omega \rangle = m_\omega f_\omega \epsilon_\mu$ & $f_\omega \approx f_{\rho} = 216$~$\mathrm{MeV}$~
\\ \hline
\multirow{2}{*}{$\bar q \gamma_\mu \gamma_5 q$} & \multirow{2}{*}{$(0^+)1^{++}$} & $\eta$               & $(0^+)0^{-+}$ & $\langle 0 | J_\mu | \eta \rangle = i p_\mu f_{\eta_q}$        & $f_{\eta_q} = 156$~$\mathrm{MeV}$~
\\ \cline{3-6} &      & $\eta^\prime$          & $(0^+)0^{-+}$ & $\langle 0 | J_\mu | \eta^\prime \rangle = i p_\mu f_{\eta^\prime_q}$ &  $f_{\eta^\prime_q} = 127$~$\mathrm{MeV}$~
\\ \hline
$\bar q \sigma_{\mu\nu} q$ & $(0^-)1^{\pm-}$ & $\omega$             & $(0^-)1^{--}$ & $\langle 0 | J_{\mu\nu} | \omega \rangle = i f^T_{\omega} (p_\mu\epsilon_\nu - p_\nu\epsilon_\mu) $ &  $f^T_{\omega} \approx f_{\rho}^T = 159$~$\mathrm{MeV}$~
\\ \hline\hline
$\bar q q$ & $(1^-)0^{++}$ & -- & $(1^-)0^{++}$ & -- & --
\\ \hline
$\bar q i\gamma_5 q$ & $(1^-)0^{-+}$ &  $\pi^0$  & $(1^-)0^{-+}$ &  $\langle 0 | J | \pi^0 \rangle = \lambda_\pi$  &  $\lambda_\pi = 256136$~$\mathrm{MeV}^2$~
\\ \hline
$\bar q \gamma_\mu q$ & $(1^+)1^{--}$ &  $\rho^0$  & $(1^+)1^{--}$ &  $\langle 0 | J_\mu | \rho^0 \rangle = m_\rho f_{\rho} \epsilon_\mu$  &  $f_{\rho} = 216$~$\mathrm{MeV}$~
\\ \hline
$\bar q \gamma_\mu \gamma_5 q$ & $(1^-)1^{++}$ & $\pi^0$  & $(1^-)0^{-+}$ & $\langle 0 | J_\mu | \pi^0 \rangle = i p_\mu f_{\pi}$  &  $f_{\pi} = 130.2$~$\mathrm{MeV}$~
\\ \hline
$\bar q \sigma_{\mu\nu} q$ & $(1^+)1^{\pm-}$ &  $\rho^0$ & $(1^+)1^{--}$   &  $\langle 0 | J_{\mu\nu} | \rho^0 \rangle = i f^T_{\rho} (p_\mu\epsilon_\nu - p_\nu\epsilon_\mu) $  &  $f_{\rho}^T = 159$~$\mathrm{MeV}$~
\\ \hline\hline
$\bar q s$ & $ 0^{+}$ & $\kappa$ & $0^{+}$ & $\langle 0 | J | \kappa \rangle = m_\kappa f_\kappa$          & $f_\kappa \sim 420 $~$\mathrm{MeV}$~
\\ \hline
$\bar q i\gamma_5 s$ & $0^{-}$ &  $K^0$  & $0^{-}$ &  $\langle 0 | J | K^0 \rangle = \lambda_K$  &  $\lambda_K = 294441$~$\mathrm{MeV}^2$~
\\ \hline
$\bar q \gamma_\mu s$ & $1^{-}$ &  $K^*$  & $1^{-}$ &  $\langle 0 | J_\mu | K^* \rangle = m_{K^*} f_{K^*} \epsilon_\mu$  &  $f_{K^*} = 226$~$\mathrm{MeV}$~
\\ \hline
$\bar q \gamma_\mu \gamma_5 s$ & $1^{+}$ & $K^0$  & $0^{-}$ & $\langle 0 | J_\mu | K^0 \rangle = i p_\mu f_{K^0}$  &  $f_{K^0} = 155.6$~$\mathrm{MeV}$~
\\ \hline
$\bar q \sigma_{\mu\nu} s$ & $1^{\pm}$ &  $K^*$ & $1^{-}$   &  $\langle 0 | J_{\mu\nu} | K^* \rangle = i f^T_{K^*} (p_\mu\epsilon_\nu - p_\nu\epsilon_\mu) $  &  $f_{K^*}^T = 185$~$\mathrm{MeV}$~
\\ \hline\hline
$\bar s s$                & $(0^+)0^{++}$                  & $f_0$ & $(0^+)0^{++}$ & $\langle 0 | J | f_0 \rangle = m_{f_0} f_{f_0}$          & $f_{f_0} = 358$~$\mathrm{MeV}$~
\\ \hline
\multirow{2}{*}{$\bar s i\gamma_5 s$ }     &  \multirow{2}{*}{$(0^+)0^{-+}$}    & $\eta$    & $(0^+)0^{-+}$ & $\langle 0 | J | \eta \rangle = \lambda_{\eta_s}$        & $\lambda_{\eta_s} = -176520$~$\mathrm{MeV}^2$~
\\ \cline{3-6} &      & $\eta^\prime$          & $(0^+)0^{-+}$ & $\langle 0 | J | \eta^\prime \rangle = \lambda_{\eta^\prime_s}$
 &  $\lambda_{\eta^\prime_s} = 187463$~$\mathrm{MeV}^2$~
\\ \hline
$\bar s \gamma_\mu s$ & $(0^-)1^{--}$                  & $\phi$             & $(0^-)1^{--}$ & $\langle 0 | J_\mu | \phi \rangle = m_\phi f_\phi \epsilon_\mu$ & $f_\phi = 233$~$\mathrm{MeV}$~
\\ \hline
\multirow{2}{*}{$\bar s \gamma_\mu \gamma_5 s$} & \multirow{2}{*}{$(0^+)1^{++}$} & $\eta$               & $(0^+)0^{-+}$ & $\langle 0 | J_\mu | \eta \rangle = i p_\mu f_{\eta_s}$        & $f_{\eta_s} = -159$~$\mathrm{MeV}$~
\\ \cline{3-6} &      & $\eta^\prime$          & $(0^+)0^{-+}$ & $\langle 0 | J_\mu | \eta^\prime \rangle = i p_\mu f_{\eta^\prime_s}$ &  $f_{\eta^\prime_s} = 200$~$\mathrm{MeV}$~
\\ \hline
$\bar s \sigma_{\mu\nu} s$ & $(0^-)1^{\pm-}$ & $\phi$             & $(0^-)1^{--}$ & $\langle 0 | J_{\mu\nu} | \phi \rangle = i f^T_{\phi} (p_\mu\epsilon_\nu - p_\nu\epsilon_\mu) $ &  $f^T_{\phi}  = 175$~$\mathrm{MeV}$~
\\ \hline\hline
\end{tabular}
\label{tab:coupling}
\end{center}
\end{table*}

\section{Fierz analysis on two-gluon glueballs}
\label{sec:Result1}

After establishing the Fierz rearrangement framework in the previous section, we now investigate the decay properties of the two-gluon glueballs. Table.~\ref{tab:result1} summarizes the relative branching ratios obtained for the two-gluon glueballs with $J^{PC}=0^{++}/0^{-+}/2^{++}/2^{-+}$. In the following, we discuss the characteristic decay patterns of these states and compare them with the available experimental information appropriately.

\begin{itemize}

\item 
For the scalar glueball, we adopt the mass of the \(f_{0}(1710)\), \(\mathrm{M}=1723~\mathrm{MeV}\)~\cite{pdg}, as the input and evaluate the relative branching ratios for eight two-body decay channels. Among these modes, the \(\pi\pi\) channel is predicted to be one of the dominant decay channels. We also find that the ratio \(\Gamma(\eta\eta)/\Gamma(\pi\pi)\) provides a representative prediction, which will be compared with the result obtained from the QCD sum rule analysis in the following section. The relative branching ratio of these two channels is found to be
\begin{eqnarray}
\frac{\mathcal{B}(\mathrm{GG};0^{++} \to \eta \eta)}{\mathcal{B}(\mathrm{GG};0^{++} \to \pi\pi)} &=& 0.42 \, ,
\end{eqnarray}
which is also consistent with the experimental measurements listed in Ref.~\cite{pdg}. This observation is consistent with the commonly discussed interpretation that the \(f_0(1710)\) might contain a substantial gluon component.

The predicted branching-ratio pattern differs significantly from that measured for the \(f_0(1500)\)~\cite{pdg,BESIII:2015rug,BESIII:2018ubj}, particularly in the \(K\bar{K}\), \(\pi\pi\), and \(\eta\eta\) channels. Within the present framework, this may indicate that the \(f_0(1500)\) cannot be described as a predominantly two-gluon scalar glueball state.

We emphasize that the present analysis is based on effective couplings~\cite{Gasser:1983yg,Ecker:1988te,Borsanyi:2010cj} obtained through Fierz rearrangements and operator--meson matrix elements. Dynamical effects associated with chiral symmetry, hadron wave-function overlaps, and final-state interactions are not explicitly included. In particular, although matrix elements proportional to \(f_\pi p^\mu\) are employed, the full dynamical consequences of PCAC~\cite{Adler:1964um,Bicudo:2003fp,Berger:2008xs,Becher:2001hv} and soft-pion constraints are beyond the scope of the present treatment. Therefore, the results should be interpreted primarily as predictions for the relative decay patterns rather than precision determinations of absolute branching fractions.

\item 
For the pseudoscalar glueball, we take the mass of the \(\eta(2370)\), with $\mathrm{M}=2377~\mathrm{MeV}$~\cite{pdg}, as input and evaluate the relative branching ratios for twenty decay modes. We find that the pseudoscalar glueball exhibits a richer decay pattern than the scalar case. In particular, the vector--vector ($VV$) channels \(\phi\phi\), \(\omega\omega\), and \(\phi\omega\) receive sizable branching fractions. It is worth noting that the relative size of the \(\omega\phi\) channel should be understood with some caution. In the present Fierz-rearrangement analysis, the channels \(\omega\omega\), \(\omega\phi\), and \(\phi\phi\) originate from the same color-singlet component in the color transformation. Therefore, at the level of color and flavor symmetry counting, these three channels are expected to be of the same order. However, this symmetry estimate does not include possible dynamical suppressions. In the ideal mixing limit, the \(\omega\phi\) final state contains both non-strange and strange hidden-flavor components. Consequently, this channel may be sensitive to \(SU(3)_F\)-breaking effects and OZI rule suppression. Thus, although the \(\omega\phi\) decay channel is not forbidden in the Fierz analysis, its physical branching fraction may be smaller than the naive symmetry-limit expectation.

We also find several important tensor--pseudoscalar decay modes, among which the \(K\bar{K}_2^*(1430)\) channel gives a relatively large contribution. These channels may serve as useful signatures for identifying a pseudoscalar two-gluon glueball. Future experimental searches in these modes could provide further information on the nature of the \(\eta(2370)\) and its possible gluonic component.

Experimentally, the \(\eta(2370)\) has been observed in the \(f_0(980)\eta^\prime\) channel~\cite{BESIII:2023wfi}. We note that this channel does not appear among the dominant decay modes in our analysis. This difference may indicate that additional nonperturbative effects, which are not included in the present constituent-level treatment, could affect the observed decay pattern. Recently, the BESIII Collaboration reported a strong suppression of the \(K \bar K^{*}(892)\) decay mode~\cite{BESIII:2026mvn}. This observation is compatible with our Fierz rearrangement analysis, where this channel does not receive a significant contribution. Although further studies including dynamical effects are still needed, the suppression of the \(K \bar K^{*}(892)\) mode provides an interesting consistency check for the possible interpretation of the \(\eta(2370)\) as a \(0^{-+}\) glueball candidate.

\item
For the tensor glueball, we use the mass of the $f_2(2340)$,
$\mathrm{M}=2346~\mathrm{MeV}$~\cite{pdg}, as input and evaluate the relative branching ratios for nine decay channels. The resulting decay pattern differs substantially from those of the scalar glueball. In particular, vector--vector ($VV$) final states are found to dominate
over pseudoscalar--pseudoscalar ($PP$) channels throughout most of the
kinematically accessible region.

A notable feature of the present analysis is that all scalar--scalar
($SS$) decay channels, such as $f_0f_0$ and
$K_0^*(700)\bar K_0^*(700)$ vanish identically within the framework adopted in this work. This behavior can be traced back to the Lorentz structure of the tensor-glueball decay amplitude. and the relevant amplitude structure is schematically given by
\begin{align}
\mathcal{M}_{SS}^{2^{++}} & \propto   (p_A^\mu + p_B^\mu) \times (p_A^\nu + p_B^\nu) \times \epsilon_{\mu\nu}(p) \\
 \nonumber & \propto p^\mu  \times p^\nu \times \epsilon_{\mu\nu}(p) =0 \, .
\end{align}

The vanishing \(SS\) amplitudes should be interpreted with some caution. In the present analysis, we describe each gluon field through its leading quark--antiquark excitation within an effective constituent picture. We then perform the Fierz rearrangements exactly at the algebraic level. However, this treatment does not include possible contributions from other nonperturbative mechanisms, higher-order transitions, or mixing effects. Such contributions may modify the decay amplitudes and could generate small corrections to the vanishing \(SS\) channels.

Nevertheless, the absence of \(SS\) contributions in all channels considered here provides an interesting indication that scalar--scalar final states may be less significant than the \(VV\) modes for a tensor two-gluon glueball. Future measurements of these decay channels could help test this decay pattern and provide further information on the gluon component of the \(f_2(2340)\).

\item 
For the pseudotensor glueball, we take the mass of the $X(2600)$, $\mathrm{M}=2618~\mathrm{MeV}$~\cite{BESIII:2022sfx} as a representative input and evaluate the relative branching ratios for twenty-one decay channels. We find that the vector--vector ($VV$) channels generally exhibit relatively large branching fractions within the present framework. We also identify several tensor--pseudoscalar ($TP$) modes, such as $K \bar K_2^*(1430)$, as potentially important decay channels. In contrast, scalar--pseudoscalar ($SP$) channels, represented by modes such as $f_0(980)\eta$, tend to have smaller branching fractions.

We note that these results rely on the effective constituent picture adopted in this work. The present analysis mainly captures the algebraic structures of the decay amplitudes, while additional dynamical effects beyond this framework may influence the detailed decay pattern. Therefore, the hierarchy among different channels should be regarded as a qualitative feature of the present approach rather than a precise prediction of the physical branching fractions.

At present, no experimentally established $J^{PC}=2^{-+}$ glueball candidate has been identified. Although the $X(2600)$ has been used here only as a representative mass input, the decay channels with relatively large branching fractions obtained in our analysis may provide useful guidance for future searches for pseudotensor glueballs. In particular, the ($VV$) and ($TP$) channels deserve further experimental investigation, which may help improve our understanding of the \(2^{-+}\) glueball sector.

\end{itemize}

\begin{table*}
    
\centering
\renewcommand{\arraystretch}{1.50}
\setlength{\tabcolsep}{1.2pt}

\caption{Relative branching ratios of two-gluon glueballs with \(J^{PC}=0^{++}/0^{-+}/2^{++}/2^{-+}\), derived via the Fierz rearrangement, where \(\sigma\) denotes \(f_0(500)\), \(\kappa\) denotes \(K_0^*(700)\), \(f_0\) denotes \(f_0(980)\), and \(K^*\) denotes \(K^*(892)\).}

\begin{minipage}[t]{0.45\textwidth}
  \centering
  \renewcommand{\arraystretch}{1.6}  
  \setlength{\tabcolsep}{15pt}       
  \vspace{0pt}
  \begin{tabular}{ c | c }
  \hline\hline
 $|\mathrm{GG};0^{++} \rangle$ & 
$\dfrac{\mathcal{B}( M_A M_B  )}{\mathcal{B}( \pi\pi )}$ 
 \\ \hline
  $\pi\pi$ & $1.00$ \\
  \hline
  $K \bar{K}$ & $1.92$ \\
  \hline
  $\kappa \bar{\kappa}$ & $0.87$ \\
  \hline
  $\eta\eta$ & $0.42$ \\
  \hline
   $\rho\rho$ & $6.22\times10^{-2}$ \\
  \hline
  $\sigma\sigma$ & $5.78\times10^{-2}$ \\
  \hline
  $\eta \eta^\prime$ & $2.22\times10^{-2}$ \\
  \hline 
  $\omega \omega$ & $9.64\times10^{-3}$ \\
  \hline\hline
  \end{tabular}
  
  \vspace{1.0cm}
  
  \begin{tabular}{ c | c }
  \hline\hline
  $|\mathrm{GG};0^{-+} \rangle$ & 
$\dfrac{\mathcal{B}( M_A M_B  )}{\mathcal{B}( f_0\eta)}$\\
  \hline
  $f_0\eta$ & $1.00$ \\
  \hline
  $\omega\phi$ & $46.38$ \\
  \hline
  $\omega\omega$ & $36.72$ \\
  \hline
  $\rho b_1(1235)$ & $13.51$ \\
  \hline
   $\phi\phi$ & $11.39$ \\
  \hline
  $K^* \bar{K}_1(1270)$ & $11.28$ \\
  \hline
  $\omega h_1(1170)$ & $5.45$ \\
  \hline
   $\eta f_0(1770)$ & $4.37$ \\
   \hline
   $\pi a_2(1320)$ & $4.14$ \\
  \hline
   $K \bar{K}_0^*(1430)$ & $3.83$ \\
   \hline
  $K \bar{K}_2^*(1430)$ & $1.88$ \\
   \hline
    $\pi a_0(1450)$ & $1.85$ \\
   \hline
  $\eta f_2(1270)$ & $1.26$ \\
   \hline
  $\kappa \bar{K} $ & $1.12$ \\
  \hline
  $f_0\eta^\prime$ & $0.31$ \\
   \hline
  $\eta f_2^\prime(1525)$ & $0.23$ \\
  \hline
  $\eta^\prime f_0(1370)$ & $0.10$ \\
  \hline
   $\eta f_0(1370)$ & $5.79\times10^{-2}$ \\
  \hline
  $\eta^\prime f_2(1270)$ & $4.47\times10^{-2}$ \\
  \hline
  $\sigma \eta$ & $6.63\times10^{-4}$ \\
  \hline
  $\sigma \eta^\prime$ & $1.31\times10^{-4}$ \\
  \hline\hline
  \end{tabular}
\end{minipage}
\hspace{-0.5cm}
\begin{minipage}[t]{0.45\textwidth}
  \centering
  \renewcommand{\arraystretch}{1.6}  
  \setlength{\tabcolsep}{15pt}       
  \vspace{0pt}

  \begin{tabular}{ c | c }
  \hline\hline
 $|\mathrm{GG};2^{++} \rangle$ & 
$\dfrac{\mathcal{B}( M_A M_B  )}{\mathcal{B}( \pi\pi )}$ 
 \\ \hline
  $\pi\pi$ & $1.00$ \\
  \hline
  $K^*\bar K^*$ & $1.16\times10^{2}$ \\
  \hline
  $\rho\rho$ & $57.69$ \\
  \hline
  $\phi\phi$ & $36.77$ \\
  \hline
  $\omega \omega$ & $24.90$ \\
  \hline
  $\omega \phi$ & $11.39$ \\
  \hline
  $K \bar{K}$ & $1.71$ \\
  \hline
  $\eta\eta$ & $1.60$ \\
  \hline
  $\eta^\prime \eta^\prime$ & $0.24$ \\
  \hline
  $\eta \eta^\prime$ & $8.46\times10^{-2}$ \\
  \hline\hline
  \end{tabular}
  
  \vspace{1.0cm}
  
  \begin{tabular}{ c | c }
  \hline\hline
  $|\mathrm{GG};2^{-+} \rangle$ & 
$\dfrac{\mathcal{B}( M_A M_B  )}{\mathcal{B}( f_0\eta)}$\\
  \hline
  $f_0\eta$ & $1.00$ \\
  \hline
  $K^*\bar K^*$ & $1.24\times10^{3}$ \\
  \hline
  $\omega\phi$ & $1.04\times10^{3}$ \\
  \hline
  $\omega\omega$ & $1.03\times10^{3}$ \\
  \hline
  $\rho\rho$ & $6.62\times10^{2}$ \\
  \hline
  $\phi\phi$ & $5.02\times10^{2}$ \\
  \hline
  $K \bar K^*$ & $60.86$ \\
  \hline
  $\pi a_2(1320)$ & $59.08$ \\
  \hline
  $K \bar{K}_2^*(1430)$ & $57.66$ \\
  \hline
  $\rho \pi$ & $34.10$ \\
  \hline
  $\eta f_2(1270)$ & $27.94$ \\
  \hline
  $\phi \eta$ & $16.11$ \\
  \hline
  $\phi \eta^\prime$ & $13.34$ \\
  \hline
  $\omega \eta$ & $13.11$ \\
  \hline
  $\eta f_2^\prime(1525)$ & $10.78$ \\
  \hline
  $\eta^\prime f_2(1270)$ & $8.77$ \\
  \hline
  $\omega \eta^\prime$ & $5.07$ \\
  \hline
  $\eta^\prime f_2^\prime(1525)$ & $2.28$ \\
  \hline
  $\kappa \bar K$ & $1.20$ \\
   \hline
  $f_0 \eta^\prime $ & $0.65$ \\
  \hline
  $\sigma \eta $ & $7.75\times10^{-4}$ \\
  \hline
  $\sigma \eta^\prime $ & $2.89\times10^{-4}$ \\
  \hline\hline
  \end{tabular}
   
\end{minipage}

\label{tab:result1}
\end{table*}

$\\$
\section{Fierz analysis on three-gluon glueballs}
\label{sec:Result2}

The Fierz rearrangement can be extended to the decay of three-gluon glueballs. In this case, each gluon field is converted into a quark--antiquark pair, and the resulting six-quark operators are reorganized into three color-singlet meson operators through two successive Fierz rearrangements. The additional rearrangement leads to a substantially richer operator structure than in the two-gluon case. To illustrate this procedure, we consider the \(0^{++}\) three-gluon glueball as an example. Starting from its interpolating current, we first rearrange one pair of quark bilinears in color and Lorentz spaces. We then perform the second Fierz rearrangement on the remaining quark fields and match the resulting operators onto three meson currents. Schematically, the two successive rearrangements take the form
\begin{align}
&  |\mathrm{GGG};0^{++}\rangle  \xleftrightarrow{~~~~~} \eta_0 = f^{ijk} g_s^3 G^{\mu \nu}_i G_{j,\nu\rho}  G_{k,\mu}^{\rho}\, 
\label{eq:expand2} \\
\nonumber & \xrightarrow{~~~~~} f^{ijk} ({M_{A}})_{\alpha}^{i} \times ({M_{B}})_{\beta}^{j} \times ({M_{C}})_{\gamma}^{k} \times \mathbb{C}_1 + \cdots \\
\nonumber & \xrightarrow{~~~~~} f^{ijk} \lambda^{i}_{ab} \lambda^{j}_{cd} \lambda^{k}_{ef} \times \bar{q}_1^{a} \gamma_{\alpha} q_2^b \times \bar{q}_3^{c} \gamma_{\beta} q_4^d  \times \bar{q}_5^{e} \gamma_{\gamma} q_6^f \times \mathbb{C}_2 + \cdots \\
\nonumber &\xrightarrow{\rm color} \delta_{af} \delta_{bc} \delta_{de} \times \bar{q}_1^{a} \gamma_{\alpha} q_2^b \times \bar{q}_3^{c} \gamma_{\beta} q_4^d \times \bar{q}_5^{e} \gamma_{\beta} q_6^f \times \mathbb{C}_3 + \cdots \\
\nonumber &\xrightarrow{\rm Fierz1} \delta_{af} \delta_{bc} \delta_{de}  \times \bar{q}_1^{a} \Gamma_{1} q_6^f  \times \bar{q}_3^{c} \gamma_{\beta} q_4^d \times\bar{q}_5^{e} \Gamma_{2} q_2^b \times \mathbb{C}_4 + \cdots \\
\nonumber &\xrightarrow{\rm Fierz2} \delta_{af} \delta_{bc} \delta_{de}  \times \bar{q}_1^{a} \Gamma_{1} q_6^f  \times \bar{q}_3^{c} \Gamma_{3} q_2^b \times\bar{q}_5^{e} \Gamma_{4} q_4^d \times \mathbb{C}_5 + \cdots \\
\nonumber &\xrightarrow{~~~~~} [\bar{q}_1^{a} \Gamma_{1} q_6^a]_A \times [\bar{q}_3^b \Gamma_{3} q_2^b]_B \times [\bar{q}_5^d \Gamma_{4} q_4^d]_C \times \mathbb{C}_5 + \cdots \, .
\end{align}
where the two rearrangements successively transform the six-quark operator into color-singlet structures associated with three final-state mesons. The detailed decomposition depends on the Lorentz structure of the three-gluon interpolating current and contains substantially more terms than the two-gluon case.

For the color rearrangement, we use the identities
\begin{eqnarray}
f^{ijk}\lambda_{i}^{ab}\lambda_{j}^{cd}\lambda_{k}^{ef} & =&
+ 2i\delta^{af}\delta^{bc}\delta^{de}-2i\delta^{ad}\delta^{be}\delta^{cf}\, ,
\\ \nonumber d^{ijk}\lambda_{i}^{ab}\lambda_{j}^{cd}\lambda_{k}^{ef} &=& +2\delta^{af}\delta^{bc}\delta^{de}+2\delta^{ad}\delta^{be}\delta^{cf}
\\  \nonumber && -\frac{4}{3}\delta^{ab}\delta^{de}\delta^{cf}
 -\frac{4}{3}\delta^{af}\delta^{cd}\delta^{be}
\\  && -\frac{4}{3}\delta^{ad}\delta^{bc}\delta^{ef}+\frac{8}{9}\delta^{ab}\delta^{cd}\delta^{ef}\, .
\end{eqnarray}

Together with the Lorentz-space Fierz relations, these identities allow us to reorganize the six-quark operators into three color-singlet meson currents. We then use the corresponding current--meson matrix elements to construct the decay amplitudes. For the two-gluon glueballs, we explicitly provide the squared decay amplitudes in Appendix~\ref{app:decay}, which allows the two-body phase-space factors to be applied directly when evaluating the relative branching ratios. The situation is more involved for three-gluon glueballs because their decays lead to three-body final states. In this case, the decay widths require an integration over the three-body phase space, usually expressed in terms of Dalitz variables, and the resulting squared amplitudes contain a considerably more complicated momentum dependence. Moreover, the unsquared decay amplitudes contain a large number of individual terms and are not particularly useful to display in full. We therefore do not list the explicit decay amplitudes or their squared forms for the three-gluon glueballs. Instead, we use the resulting amplitudes to evaluate the corresponding phase-space integrals and present the relative branching ratios directly.

We apply this framework to the three-gluon glueballs with \(J^{PC}=0^{++}\) and \(1^{+-}\) and summarize their resulting branching ratios in Table~\ref{tab:result2}. For the \(0^{++}\) state, we take \(\mathrm{M}=4210~\mathrm{MeV}\), as estimated in Ref.~\cite{Chen:2021bck}, and evaluate the relative branching ratios for thirty-seven decay channels. Among the channels considered, the \(\pi\pi\omega\) and \(K\bar K\phi\) modes have relatively large branching fractions. For the \(1^{+-}\) state, we use \(\mathrm{M}=3190~\mathrm{MeV}\) from Ref.~\cite{Chen:2021bck} and obtain relative branching ratios for twenty-nine channels. The \(\pi\pi\omega\), \(\pi\pi\phi\), and \(K\bar K\phi\) channels are among the more prominent modes.

\begin{table*}
    
\centering
\renewcommand{\arraystretch}{1.50}
\setlength{\tabcolsep}{1.2pt}

\caption{Relative branching ratios of three-gluon glueballs with \(J^{PC}=0^{++}/1^{+-}\), derived via the Fierz rearrangement, where \(\sigma\) denotes \(f_0(500)\), \(\kappa\) denotes \(K_0^*(700)\), \(f_0\) denotes \(f_0(980)\), and \(K^*\) denotes \(K^*(892)\).}

\begin{minipage}[t]{0.45\textwidth}
  \centering
   \renewcommand{\arraystretch}{1.6}  
  \setlength{\tabcolsep}{15pt}       
  
  \vspace{0pt}
   \begin{tabular}{ c | c }
  \hline\hline
  $|\mathrm{GGG};0^{++} \rangle$ & 
$\dfrac{\mathcal{B}( M_A M_B M_C )}{\mathcal{B}( \pi\pi\omega )}$ \\
\hline  
 $\pi\pi\omega$ & $1.00$ \\
  \hline
  $K\bar{K}\phi$ & $0.75$ \\
   \hline
   $K\bar{K}\rho$ & $0.69$ \\
   \hline
  $K\bar{K}\omega$ & $0.35$ \\
   \hline
  $\rho\rho\omega$ & $0.35$ \\
   \hline
  $\kappa\bar{\kappa}\rho$ & $0.26$ \\
  \hline
  $\kappa\bar{\kappa}\phi$ & $0.25$ \\   
  \hline
  $K^*\bar{K}^*\rho$ & $0.24$ \\
   \hline
  $K^*\bar{K}^*\phi$ & $0.21$ \\
   \hline
  $\pi\pi f_2(1270)$ & $0.14$ \\
  \hline
   $\kappa\bar{\kappa}\omega$ & $0.13$ \\
  \hline
  $K^*\bar{K}^*\omega$ & $0.12$ \\
   \hline
  $K\bar{K}a_2(1320)$ & $7.59\times10^{-2}$ \\
   \hline
  $K\bar{K} f_2(1270)$ & $4.72\times10^{-2}$ \\
  \hline
  $K\bar{K} f_2^\prime(1525)$ & $4.34\times10^{-2}$ \\
  \hline
  $\eta \eta \phi$ & $2.24\times10^{-2}$ \\
  \hline 
   $\eta^\prime \eta^\prime \phi$ & $1.78\times10^{-2}$ \\
  \hline
  $\kappa \bar{\kappa} a_2(1320)$ & $1.78\times10^{-2}$ \\
  \hline
   $f_0f_0\phi$ & $1.76\times10^{-2}$ \\
   \hline
  $\kappa \bar{\kappa} f_2(1270)$ & $1.12\times10^{-2}$ \\
  \hline
  $\kappa \bar{\kappa} f_2^\prime(1525)$ & $9.62\times10^{-3}$ \\
  \hline
  $\eta\eta\omega$ & $8.82\times10^{-3}$ \\
  \hline
  $\phi\phi\phi$ & $6.88\times10^{-3}$ \\
  \hline
  $\omega\omega\omega$ & $5.64\times10^{-3}$ \\
  \hline
 $\kappa \bar{\kappa} f_0$ & $5.40\times10^{-3}$ \\
  \hline
   $\sigma\sigma\omega$ & $4.76\times10^{-3}$ \\
  \hline
   $\eta\eta^\prime f_2^\prime(1525)$ & $1.68\times10^{-3}$ \\
  \hline
  $\eta\eta f_2(1270)$ & $1.47\times10^{-3}$ \\
  \hline
  $\eta^\prime\eta^\prime \omega$ & $1.43\times10^{-3}$ \\
  \hline
  $\eta\eta f_2^\prime(1525)$ & $1.35\times10^{-3}$ \\
  \hline
  $\eta\eta^\prime f_2(1270)$ & $9.02\times10^{-4}$ \\
  \hline
  $\kappa \bar{\kappa} \sigma$ & $5.62\times10^{-4}$ \\
  \hline
   $\sigma \sigma f_2(1270)$ & $4.85\times10^{-4}$ \\
   \hline
  $\eta^\prime\eta^\prime f_2^\prime(1525)$ & $3.59\times10^{-4}$ \\
  \hline
  $f_0f_0f_2^\prime(1525)$ & $2.59\times10^{-4}$ \\
  \hline
  $f_0f_0f_0$ & $2.24\times10^{-4}$ \\
  \hline
   $\eta^\prime\eta^\prime f_2(1270)$ & $1.10\times10^{-4}$ \\
  \hline
   $\sigma \sigma \sigma $ & $4.67\times10^{-8}$ \\
  \hline\hline
  \end{tabular}
\end{minipage}
\hspace{-0.5cm}
\begin{minipage}[t]{0.45\textwidth}
  \centering
   \renewcommand{\arraystretch}{1.6}  
  \setlength{\tabcolsep}{15pt}       
  
  \vspace{0pt}
  \begin{tabular}{ c | c }
  \hline\hline
  $|\mathrm{GGG};1^{+-} \rangle$ & 
$\dfrac{\mathcal{B}( M_A M_B M_C )}{\mathcal{B}( \pi\pi\omega )}$ \\
  \hline
   $\pi\pi\omega$ & $1.00$ \\
  \hline
  $\pi\pi\phi$ & $1.49$ \\
  \hline
  $K\bar{K} \omega$ & $0.83$ \\
  \hline
   $K\bar{K}\phi$ & $0.27$ \\
  \hline
  $\kappa \bar{\kappa} \omega$ & $0.21$ \\
  \hline
   $\eta\eta\omega$ & $0.16$ \\
  \hline
  $K\bar{K}\rho$ & $0.15$ \\
  \hline
  $\kappa \bar{\kappa} \phi$ & $0.11$ \\
  \hline
  $\eta\eta\phi$ & $7.62\times10^{-2}$ \\
  \hline
   $\sigma\sigma\omega$ & $4.69\times10^{-2}$ \\
  \hline
  $\sigma\sigma\phi$ & $2.88\times10^{-2}$ \\
  \hline
  $\eta^\prime \eta^\prime \omega$ & $4.62\times10^{-3}$ \\
  \hline
  $K\bar{K}f_0$ & $3.65\times10^{-3}$ \\
  \hline
   $\eta \eta^\prime \omega$ & $3.51\times10^{-3}$ \\
  \hline
  $f_0f_0\omega$ & $5.38\times10^{-4}$ \\
  \hline
   $\eta \eta^\prime \phi$ & $4.95\times10^{-4}$ \\
  \hline
  $\eta^\prime \eta^\prime \phi$ & $4.12\times10^{-4}$ \\
  \hline
  $\eta \eta f_0$ & $8.77\times10^{-5}$ \\
  \hline
  $\eta \eta^\prime f_0$ & $4.58\times10^{-5}$ \\
  \hline
   $f_0f_0\phi$ & $3.61\times10^{-5}$ \\
  \hline
  $\pi\pi\sigma$ & $9.35\times10^{-6}$ \\
  \hline 
  $K\bar{K}\sigma$ & $2.80\times10^{-6}$ \\
  \hline
  $\eta^\prime \eta^\prime f_0$ & $1.27\times10^{-6}$ \\
  \hline
  $\eta \eta \sigma$ & $8.05\times10^{-8}$ \\
  \hline
  $\eta \eta^\prime \sigma$ & $3.40\times10^{-8}$ \\
  \hline
  $\eta^\prime \eta^\prime \sigma$ & $2.38\times10^{-9}$ \\
  \hline
  $K\bar{K}a_1(1260)$ & $1.62\times10^{-15}$ \\
  \hline
   $K\bar{K}f_1(1420)$ & $7.18\times10^{-16}$ \\
  \hline
  $K\bar{K}f_1(1285)$ & $6.25\times10^{-16}$ \\
  \hline
  $\pi\pi f_1(1285)$ & $3.58\times10^{-16}$ \\
  \hline\hline
  \end{tabular}
\end{minipage}

\label{tab:result2}
\end{table*}

$\\$
\section{QCD sum rule analysis for two-gluon scalar glueball decay process}
\label{sec:Sum rule}

In this section, we investigate the two-body decays of the scalar glueball with $J^{PC}=0^{++}$ within the framework of QCD sum rules using three-point correlation functions. As an independent approach, the present calculation provides a useful comparison with the Fierz rearrangement results obtained in this work. The decay widths of eight possible two-body channels are evaluated, and the numerical results are summarized in Table~\ref{tab:QCD}.

As an illustration of the calculation procedure, we take the decay mode

\[|\mathrm{GG};0^{++}\rangle \rightarrow \pi\pi\] as an example to study the normal decay process through the three-point correlation function. The decay property is studied through the three-point correlation function

\begin{align}
T(p,k,q)=\int d^4x d^4y e^{ikx}e^{iqy}
\langle 0|\mathbb{T}[J_5^\pi(x)J_5^\pi(y)J_{0^{++}}^\dagger(0)]|0\rangle ,
\end{align}

where \(p_\mu\), \(k_\mu\), and \(q_\mu\) denote the four-momenta of the scalar glueball \(\mathrm{GG}_{0^{++}}\) and the two final-state pions, respectively. The interpolating current \(J_{0^{++}}\)has been defined previously. The pseudoscalar current \(J_\pi\) represents the pseudoscalar current coupling to the pion state with the decay constant $\lambda_\pi$.

At the phenomenological side, the three point correlation function is saturated by inserting the intermediate physical states. The contribution from the lowest scalar glueball and pion states can be expressed as

\begin{eqnarray}
T^{\rm PH}(p,k,q)=
\frac{g_{\pi\pi} f_G \lambda_\pi^2}
{(p^2-m_G^2)(k^2-m_\pi^2)(q^2-m_\pi^2)}
+\cdots .
\end{eqnarray}

where the coupling constant \(g_{\pi\pi}\) is introduced through the effective interaction Lagrangian

\begin{align}
\mathcal{L}_{eff}=\frac{1}{2} g_{\pi\pi}\mathrm{GG}_{0^{++}}(\pi^0\pi^0+ 2\pi^+\pi^-) .
\end{align}

The above expression describes the transition of the scalar glueball into two pseudoscalar mesons at the hadronic level.

On the QCD side, the correlation function is evaluated by means of the operator product expansion (OPE), where the short-distance contributions are expanded in terms of local operators and the corresponding vacuum condensates. After performing the Fourier transformation and applying the adopted kinematic conditions, the two final-state mesons are taken to have equal momentum, \(k^{2}=q^{2}\). The QCD representation of the correlation function can then be written as

\begin{align}
T^{\rm OPE}(p,k,q)=\frac{ \langle g_s^2 GG \rangle^2}{48}\times \frac{\pi^2}{k^2} ,
\end{align}

In numerical analysis, the double-gluon condensate is taken as \(\langle g_s^2 GG \rangle = 0.88 \pm 0.15 ~\mathrm{GeV}^4\)~\cite{Navarra:2007yw,Narison:2010cg,Lu:2023pcg}. For the scalar glueball, we adopt \(m_G=1723^{+7}_{-6}~\mathrm{MeV}\) corresponding to the mass of the \(f_0(1710)\)~\cite{pdg} candidate state as the input parameter.

By matching the phenomenological representation with the QCD calculation, the coupling constant \(g_{\pi\pi}\) can be extracted. After applying once Borel transformation to suppress the contributions we arrive at

\begin{eqnarray}
-g_{\pi\pi}\frac{f_G\lambda_\pi^2 e^{\frac{-m_\pi^2}{T^2}}}{m_G^2 T^2}= -\frac{ \langle g_s^2 GG \rangle^2 \pi^2}{48} .
\end{eqnarray}

The decay width of the scalar glueball into two pions is determined by

\begin{eqnarray}
\Gamma(\mathrm{GG}_{0^{++}}\rightarrow\pi\pi)
=
\frac{3}{2}\times\frac{g_{\pi\pi}^{\,2}}{16\pi m_G}
\sqrt{1-\frac{4m_\pi^2}{m_G^2}},
\end{eqnarray}

Using the standard QCD sum rule parameters and numerical inputs, the coupling constant and decay width are obtained as

\begin{eqnarray}
g_{\pi\pi} &=& 1.37 ^{+0.57}_{-0.46} ~ \mathrm{GeV} ,
\\
\Gamma_{G_{0^{++}}\rightarrow\pi\pi} &=&  32.25^{+29.30}_{-18.01}~ \mathrm{MeV}.
\end{eqnarray}

\begin{table}[htbp]
\centering
\caption{Partial decay widths of the scalar glueball $\mathrm{GG}_{0^{++}}$ in units of MeV. We simply sum over the partial decay widths to obtain the total decay widths, as listed in the last row. Here we make \(\sigma\) denotes \(f_0(500)\), \(\kappa\) denotes \(K_0^*(700)\).}

\renewcommand{\arraystretch}{1.5}  
\setlength{\tabcolsep}{18pt} 
\begin{tabular}{c | c}

\hline\hline
Channel & Decay width $~$ ($\mathrm{MeV}$) \\
\hline
$\mathrm{GG}_{0^{++}}\rightarrow \pi\pi$ 
& $ 32.25^{+29.30}_{-18.01} $  \\

$\mathrm{GG}_{0^{++}}\rightarrow K\bar{K}$ 
& $ 28.39^{+25.78}_{-15.84} $ \\

$\mathrm{GG}_{0^{++}}\rightarrow \eta\eta$ 
& $ 14.25^{+12.94}_{-7.95} $ \\

$\mathrm{GG}_{0^{++}}\rightarrow \eta\eta'$ 
& $ 1.01^{+0.91}_{-0.56} $ \\

$\mathrm{GG}_{0^{++}}\rightarrow \sigma\sigma$ 
& $ 6.34^{+5.75}_{-3.52} $ \\

$\mathrm{GG}_{0^{++}}\rightarrow \kappa\bar{\kappa}$ 
& $ 24.99^{+22.69}_{-13.98} $ \\

$\mathrm{GG}_{0^{++}}\rightarrow \rho\rho$ 
& $ 0 $ \\

$\mathrm{GG}_{0^{++}}\rightarrow \omega\omega$ 
& $ 0 $ \\

\hline
Sum
 & $ 107^{+94}_{-56} $ 
\\\hline\hline

\end{tabular}

\label{tab:QCD}
\end{table}

It should be emphasized that the present QCD sum rule analysis is intended as a leading-order study. In the current calculation, we retain only the dominant Lorentz structure and do not include subleading contributions or corrections induced by higher-order interactions. Consequently, the present results should be regarded as a first estimate of the decay properties rather than a complete description. Within this approximation, we find vanishing contributions to the \(\rho\rho\) and \(\omega\omega\) channels, indicating that these decay modes are mainly generated by subleading effects beyond the present treatment. Their decay widths are therefore expected to be relatively small.

Despite these limitations, we observe an encouraging agreement between the QCD sum rule calculation and the Fierz rearrangement analysis for the dominant decay channels. For example, the predicted ratio  \(\Gamma(\eta\eta)/\Gamma(\pi\pi) \sim  0.44 \) is close to that obtained from the Fierz approach. A similar level of consistency between the two methods has also been reported in Ref.~\cite{Su:2025bhv}. Although the present QCD sum rule analysis remains approximate, these observations suggest that the Fierz rearrangement method captures the dominant features of the two-body decay pattern of the scalar glueball.

\section{Conclusion}
\label{sec:Summary}
In this work, we have systematically studied the decay properties of two- and three-gluon glueballs using the Fierz rearrangement method. For the two-gluon systems, we considered the \(J^{PC}=0^{++}\), \(0^{-+}\), \(2^{++}\), and \(2^{-+}\) states and obtained their relative branching ratios for a broad set of decay channels. We also performed an independent QCD sum rule analysis of the \(0^{++}\) two-gluon glueball using three-point correlation functions. The consistency between the two approaches for the dominant scalar decay channels provides a useful check of the Fierz-based analysis.

For the scalar and pseudoscalar glueballs, our results support a sizable gluon component in the \(f_0(1710)\) and favor the interpretation of the \(\eta(2370)\) as a \(0^{-+}\) glueball candidate. In particular, the decay pattern obtained for the \(\eta(2370)\) is consistent with recent experimental observations which the \(K \bar K^{*}(892)\) mode is found to be strongly suppressed experimentally~\cite{BESIII:2026mvn}. This agreement strengthens the case for its \(0^{-+}\) glueball interpretation. For the tensor glueball, we find that the decay pattern is dominated by the vector--vector \((VV)\) channels, with \(K^{*}(892)\bar K^{*}(892)\) being particularly promising for experimental searches. Further measurements of this channel may provide additional information on the nature of tensor glueball candidates.

We have also extended the analysis to three-gluon glueballs with \(J^{PC}=0^{++}\) and \(1^{+-}\). The three-gluon systems require two successive Fierz rearrangements and lead to three-body decay amplitudes. We obtain relative branching ratios for a range of three-body channels and identify the \(\pi\pi\omega\) and \(K\bar K\phi\) modes as promising channels for the \(0^{++}\) state, while the \(\pi\pi\omega\), \(\pi\pi\phi\), and \(K\bar K\phi\) modes are among the more prominent channels for the \(1^{+-}\) state. These results may provide useful guidance for future experimental searches for two- and three-gluon glueballs. We hope that further measurements of the suggested channels will help clarify the nature of these states and test the decay patterns obtained in the present framework.

\section*{Acknowledgments}

We thank Niu Su for useful discussions. This project is supported by the National Natural Science Foundation of China under Grant No.~12075019, the Jiangsu Provincial Double-Innovation Program under Grant No.~JSSCRC2021488, the SEU Innovation Capability Enhancement Plan for Doctoral Students No.~CXJHSEU25139,
and the Fundamental Research Funds for the Central Universities.

\appendix

\begin{widetext}
\section{Full Fierz rearrangement result of the two-gluon scalar glueball}
\label{app:Fierz}

In this appendix, we give further details of the derivation of Eq.~(\ref{eq:expand1}) and provide the complete Fierz result for the quark-flavor structure \(\bar u d\). We retain the intermediate steps to clarify how the color and Lorentz rearrangements lead to the final meson--meson operators.

\begin{align}
|\mathrm{GG};0^{++}\rangle &\xleftrightarrow{~~~~~} J_0 = G^{\mu \nu}_i \times G_{\mu \nu}^i \times g_s^2
 \\
\nonumber &\xrightarrow{~~~~~} \mathbb{C}_1 \times (\partial^{\mu} A^{\nu}_{i} - \partial^{\nu} A^{\mu}_{i} + g_sf_{ijk} A^{\mu}_{j}A^{\nu}_{k} ) \times (\partial_{\mu} A_{\nu}^{i} - \partial_{\nu} A_{\mu}^{i} + g_sf^{ijk} A_{\mu}^{j}A_{\nu}^{k} )  \\
\nonumber &\xrightarrow{~~~~~} \mathbb{C}_2  \times \lambda^{i}_{ab} \lambda^{i}_{cd} \times ( \partial^{\mu}\bar{q}_1^{a} \gamma^{\nu} q_2^b + \bar{q}_1^{a} \gamma^{\nu} \partial^{\mu}q_2^b)\times ( \partial_{\mu}\bar{q}_3^{c} \gamma_{\nu} q_4^d + \bar{q}_3^{c} \gamma_{\nu} \partial_{\mu}q_4^d)  + \cdots \\
\nonumber &\xrightarrow{\rm color} \mathbb{C}_3 \times \delta_{ad} \delta_{cb} \times  \partial^{\mu}\bar{q}_1^{a} \gamma^{\nu} q_2^b\times  \partial_{\mu}\bar{q}_3^{c} \gamma_{\nu} q_4^d  + \cdots \\
\nonumber &\xrightarrow{\rm Fierz} \mathbb{C}_4 \times \delta_{ad} \delta_{cb} \times (-\partial^{\mu}\bar{q}_1^{a}  q_4^d \times \partial_{\mu}\bar{q}_3^{c}  q_2^b + \frac{1}{2}\partial^{\mu}\bar{q}_1^{a} \gamma^{\nu} q_4^d \times \partial_{\mu}\bar{q}_3^{c} \gamma_{\nu} q_2^b 
\\ \nonumber & + \frac{1}{2}\partial^{\mu}\bar{q}_1^{a} \gamma^{\nu}\gamma_5 q_4^d \times \partial_{\mu}\bar{q}_3^{c} \gamma_{\nu}\gamma_5 q_2^b + \partial^{\mu}\bar{q}_1^{a} \gamma_5 q_4^d \times \partial_{\mu}\bar{q}_3^{c} \gamma_5 q_2^b) + \cdots \\
\nonumber &\xrightarrow{~~~~~} \mathbb{C}_5 \times -( \partial^{\mu}[\bar{q}_1^{a} q_4^a]_A - [\bar{q}_1^{a} \overleftrightarrow{\partial}^\mu q_4^a]_A ) \times ( \partial_{\mu}[\bar{q}_3^{b} q_2^b]_B - [\bar{q}_3^{b} \overleftrightarrow{\partial}_\mu q_2^b]_B )  + \cdots \\
\nonumber &\xrightarrow{~~~~~} +\frac{1}{4}[\bar{u}_1^a \gamma_{\mu}\gamma_5 d_4^a]_A [\bar{d}_3^b \gamma_{\mu}\gamma_5 u_2^b]_B \times(q_A^{\nu}+q_B^{\nu})^2  +\frac{1}{2}[\bar{u}_1^a \gamma_{\mu}\gamma_5 d_4^a]_A [\bar{d}_3^b \gamma_{\nu}\gamma_5 u_2^b]_B \times (q_A^{\mu}+q_B^{\mu})(q_A^{\nu}+q_B^{\nu})
\\ \nonumber & +\frac{3}{4}[\bar{u}_1^a \gamma_5 d_4^a]_A [\bar{d}_3^b \gamma_5 u_2^b]_B\times (q_A^2+q_B^2+2q_A \cdot  q_B) +\frac{1}{4}[\bar{u}_1^a \gamma_{\mu} d_4^a]_A [\bar{d}_3^b \gamma_{\mu} u_2^b]_B \times(q_A^{\nu}+q_B^{\nu})^2
\\ \nonumber & +\frac{1}{2}[\bar{u}_1^a \gamma_{\mu} d_4^a]_A [\bar{d}_3^b \gamma_{\nu} u_2^b]_B \times (q_A^{\mu}+q_B^{\mu})(q_A^{\nu}+q_B^{\nu}) -\frac{1}{2}[\bar{u}_1^a \sigma_{\mu\alpha} d_4^a]_A [\bar{d}_3^b \sigma_{\nu\alpha} u_2^b]_B \times (q_A^{\mu}+q_B^{\mu})(q_a^{\nu}+q_b^{\nu})
\\ \nonumber & +\frac{1}{8}[\bar{u}_1^a \sigma_{\mu\nu} d_4^a]_A [\bar{d}_3^b \sigma_{\mu\nu} u_2^b]_B \times (q_A^2+q_B^2+2q_A \cdot  q_B) -
\frac{3}{4} [\bar{u}_1^a  d_4^a]_A [\bar{d}_3^b  u_2^b]_B \times (q_A^2+q_B^2+2 q_A \cdot  q_B)
\\ \nonumber & -\frac{3}{2} [\bar{u}_1^a \overleftrightarrow{\partial}^\mu d_4^a]_A [\bar{d}_3^b \overleftrightarrow{\partial}_\mu u_2^b]_B + \frac{3}{2} [\bar{u}_1^a \overleftrightarrow{\partial}^\mu \gamma_5 d_4^a]_A [\bar{d}_3^b \overleftrightarrow{\partial}_\mu \gamma_5 u_2^b]_B  +\frac{1}{2} [\bar{u}_1^a \overleftrightarrow{\partial}^\mu \gamma_{\mu} d_4^a]_A [\bar{d}_3^b \overleftrightarrow{\partial}_\mu \gamma^{\mu} u_2^b]_B 
\\ \nonumber & +\frac{1}{2} [\bar{u}_1^a \overleftrightarrow{\partial}^\mu \gamma_{\nu} d_4^a]_A [\bar{d}_3^b \overleftrightarrow{\partial}_\mu \gamma^{\nu} u_2^b]_B  +\frac{1}{2} [\bar{u}_1^a \overleftrightarrow{\partial}^\mu \gamma_{\nu} d_4^a]_A [\bar{d}_3^b \overleftrightarrow{\partial}_\nu \gamma^{\mu} u_2^b]_B +\frac{1}{2} [\bar{u}_1^a \overleftrightarrow{\partial}^\mu \gamma_{\mu} \gamma_5 d_4^a]_A [\bar{d}_3^b \overleftrightarrow{\partial}_\mu \gamma^{\mu} \gamma_5 u_2^b]_B 
\\ \nonumber & + \frac{1}{2} [\bar{u}_1^a \overleftrightarrow{\partial}^\mu \gamma_{\nu} \gamma_5 d_4^a]_A [\bar{d}_3^b \overleftrightarrow{\partial}_\mu \gamma^{\nu} \gamma_5 u_2^b]_B + \frac{1}{2} [\bar{u}_1^a \overleftrightarrow{\partial}^\mu \gamma_{\nu} \gamma_5 d_4^a]_A [\bar{d}_3^b \overleftrightarrow{\partial}_\nu \gamma^{\mu} \gamma_5 u_2^b]_B - \frac{1}{2} [\bar{u}_1^a \overleftrightarrow{\partial}^\mu \sigma_{\mu\alpha} d_4^a]_A [\bar{d}_3^b \overleftrightarrow{\partial}_\nu \sigma^{\nu\alpha}  u_2^b]_B
\\ \nonumber &  - \frac{1}{2} [\bar{u}_1^a \overleftrightarrow{\partial}^\mu \sigma_{\nu\alpha} d_4^a]_A [\bar{d}_3^b \overleftrightarrow{\partial}_\nu \sigma^{\mu\alpha} \gamma_5 u_2^b]_B + \frac{1}{4} [\bar{u}_1^a \overleftrightarrow{\partial}^\mu \sigma_{\nu\alpha} d_4^a]_A [\bar{d}_3^b \overleftrightarrow{\partial}_\mu \sigma^{\nu\alpha} \gamma_5 u_2^b]_B\, .
\end{align}

The global factors $\mathbb{C}_1$, $\mathbb{C}_2$, $\mathbb{C}_3$, $\mathbb{C}_4$ and $\mathbb{C}_5$ appearing at different stages of the derivation are related to one another and should not be regarded as independent parameters. Their absolute values, however, cannot be fixed within the present treatment. In our construction, we represent the gluon field through an effective vector-current coupling to a quark--antiquark pair. This prescription specifies the algebraic structure of the transition but does not determine its overall dynamical strength. We therefore leave the common normalization undetermined. Since our analysis compares relative branching ratios, this overall factor cancels in the ratios and does not affect the results discussed in the main text.

For clarity, we use ellipses in the intermediate equations to denote terms that arise at the corresponding stage but are not needed to follow the rearrangement. We retain the complete set of terms in the final Fierz result, where all relevant color and Lorentz structures are displayed explicitly.

\end{widetext}

\begin{widetext}
\section{Squared decay amplitudes for two-gluon glueballs decay channels in Fierz analysis}
\label{app:decay}

In this appendix, we present the squared decay amplitudes for the two-body decay channels of the two-gluon glueballs with
\(J^{PC}=0^{++}\), \(0^{-+}\), \(2^{++}\), and \(2^{-+}\).

The analytical expressions for the decay amplitudes themselves are generally rather lengthy due to the Lorentz structures associated with the glueball currents and the polarization tensors of the final state mesons. By contrast, after summing over the polarizations and applying the relevant kinematic relations, the corresponding squared amplitudes can be expressed in a much more compact form. Since the decay widths depend directly on the squared amplitudes, we list only these expressions below.

For definiteness, we consider final states with isospin \(I=0\) with the quark content is represented by
\[
\frac{u\bar u+d\bar d}{\sqrt{2}}.
\]
Such flavor configurations are relevant for the isoscalar mesons considered in the present work and allow all corresponding decay channels to be treated in a unified manner. The expressions for other flavor assignments can be obtained straightforwardly by replacing the relevant quark flavors.

Throughout the derivation, all external particles are taken to be on shell, and the standard kinematic identities
\[
p^2=M^2,\qquad
p_1^2=m_1^2,\qquad
p_2^2=m_2^2,
\]
together with
\[
p=p_1+p_2,\qquad
p\!\cdot\!p_1=\frac{M^2+m_1^2-m_2^2}{2},\qquad
p\!\cdot\!p_2=\frac{M^2+m_2^2-m_1^2}{2},
\]
and
\[
p_1\!\cdot\!p_2=\frac{M^2-m_1^2-m_2^2}{2},
\]
have been employed to simplify the analytical expressions.

The squared decay amplitudes for the individual channels are listed below.

\begin{itemize}

\item The squared decay amplitudes of $| \mathrm{GG};0^{++}\rangle \to \sigma \sigma$ can be expressed as
\begin{eqnarray}
 \mathcal{|M|}_{\sigma \sigma}^2 \propto \frac{9}{32} f_\sigma^4 m_\sigma^4 m_G^4 \, ,
\end{eqnarray}

\item The squared decay amplitudes of $| \mathrm{GG};0^{++}\rangle \to \eta \eta$ can be expressed as
\begin{eqnarray}
\mathcal{|M|}_{\eta \eta}^2 &\propto& \frac{1}{32} m_G^4 (3\lambda^2_{\eta_q} + f^2_{\eta_q}(m^2_G-m^2_\eta))^2 \, .
\end{eqnarray}

\item The squared decay amplitudes of $| \mathrm{GG};0^{++}\rangle \to \omega \omega$ can be expressed as
\begin{eqnarray}
\mathcal{|M|}_{\omega \omega}^2 &\propto& \frac{1}{288} [ 27 {f^T_{\omega}}^4 m_\omega^4 m_G^4 +6 f_\omega^2 {f^T_{\omega}}^2 m_\omega^2 m_G^2(32 m_\omega^4 -35 m_\omega^2 m_G^2 + 3m_G^4) 
\\ \nonumber && + f_\omega^4(3072 m_\omega^8-1088 m_\omega^6 m_G^2 + 227 m_\omega^4 m_G^4 -54 m_\omega^2 m_G^6 +9 m_G^8) ]\, .
\end{eqnarray}

\item The squared decay amplitudes of $| \mathrm{GG};0^{-+}\rangle \to \sigma \eta$ can be expressed as
\begin{eqnarray}
 \mathcal{|M|}_{\sigma \eta}^2 \propto \frac{1}{4} (m_u+m_d)^2 f^2_{\eta_q} f_\sigma^2 m_\sigma^2 (3m_\eta^2 + m_\sigma^2 -m_G^2)^2 \, ,
\end{eqnarray}

\item The squared decay amplitudes of $| \mathrm{GG};0^{-+}\rangle \to \eta f_0(1370)$ can be expressed as
\begin{eqnarray}
 \mathcal{|M|}_{\eta f_0}^2 \propto \frac{1}{4} \lambda^2_{\eta_q}  f_{f_0}^2  (m_\eta^2 + 3m_{f_0}^2 -m_G^2)^2 \, .
\end{eqnarray}

\item The squared decay amplitudes of $| \mathrm{GG};0^{-+}\rangle \to \eta f_2(1270)$ can be expressed as
\begin{eqnarray}
 \mathcal{|M|}_{\eta f_2}^2 \propto \frac{1}{24 m_{f_2}^4} f^2_{\eta_q}  f_{f_2}^2  (m_\eta^4 + (m_{f_2}^2-m_G^2)^2 -2 m_\eta^2 (m_{f_2}^2+m_G^2))^2 \, .
\end{eqnarray}

\item The squared decay amplitudes of $| \mathrm{GG};0^{-+}\rangle \to \omega h_1(1170)$ can be expressed as
\begin{eqnarray}
 \mathcal{|M|}_{\omega h_1}^2 \propto \frac{1}{4 m_{h_1}^2} {f^T_{\omega}}^2  f_{h_1}^2  (2m_{h_1}^6 + m_{h_1}^4(21m_\omega^2-4m_G^2) +(m_\omega^3-m_\omega m_G^2)^2+2 m_{h_1}^2 (12m_\omega^4 -9 m_\omega^2 m_G^2 +m_G^4) ) \, .
\end{eqnarray}

\item The squared decay amplitudes of $| \mathrm{GG};0^{-+}\rangle \to \omega \omega$ can be expressed as
\begin{eqnarray}
 \mathcal{|M|}_{\omega \omega}^2 \propto \frac{16}{9} f_\omega^4 m_\omega^4 m_G^2 (m_G^2-4 m_\omega^2) \, .
\end{eqnarray}

\item The squared decay amplitudes of $| \mathrm{GG};2^{++}\rangle \to \eta \eta$ can be expressed as
\begin{eqnarray}
\mathcal{|M|}_{\eta \eta}^2 &\propto& \frac{25}{3072} f^4_{\eta_q} m_G^8 (m^2_G - 4 m^2_\eta)^2 \, .
\end{eqnarray}

\item The squared decay amplitudes of $| \mathrm{GG};2^{++}\rangle \to \omega \omega$ can be expressed as
\begin{eqnarray}
\mathcal{|M|}_{\omega \omega}^2 &\propto& \frac{25}{9216} [ 36 {f^T_{\omega}}^4 m_G^8 (6 m_\omega^4 - 3 m_\omega^2 m_G^2 + m_G^4) -48 f_\omega^2 {f^T_{\omega}}^2 m_\omega^2 m_G^6(176 m_\omega^4 -123 m_\omega^2 m_G^2 + 11 m_G^4) 
\\ \nonumber && + f_\omega^4 m_G^4 (120832 m_\omega^8- 43776 m_\omega^6 m_G^2 + 5160 m_\omega^4 m_G^4 + 36 m_\omega^2 m_G^6 + 3 m_G^8) ]\, .
\end{eqnarray}

\item The squared decay amplitudes of $| \mathrm{GG};2^{-+}\rangle \to \sigma \eta$ can be expressed as
\begin{eqnarray}
 \mathcal{|M|}_{\sigma \eta}^2 \propto \frac{1}{24} (m_u+m_d)^2 f^2_{\eta_q} f_\sigma^2 m_\sigma^2 (m_\eta^4 + (m_\sigma^2 -m_G^2)^2-2m_\eta^2 (m_\sigma^2 + m_G^2))^2 \, ,
\end{eqnarray}

\item The squared decay amplitudes of $| \mathrm{GG};2^{-+}\rangle \to \omega \eta$ can be expressed as
\begin{eqnarray}
 \mathcal{|M|}_{\omega \eta}^2 &\propto& \frac{1}{1536} f^2_{\eta_q} f_\omega^2 m_G^4 [ (m_\eta^8 - 4 m_\eta^6 (m_\omega^2 +m_G^2) + (m_\omega^2 -m_G^2)^2 (m_\omega^4 +8 m_\omega^2 m_G^2 + m_G^4) 
 \\ \nonumber && + 2 m_\eta^4 (3m_\omega^4 +7 m_\omega^2 m_G^2 +3m_G^4) -4 m_\eta^2(m_\omega^6+ 4 m_\omega^4 m_G^2 +4 m_\omega^2 m_G^4 +m_G^6) ] \, ,
\end{eqnarray}

\item The squared decay amplitudes of $| \mathrm{GG};2^{-+}\rangle \to \eta f_2(1270)$ can be expressed as
\begin{eqnarray}
 \mathcal{|M|}_{\eta f_2}^2 &\propto& \frac{f^2_{\eta_q}  f_{f_2}^2}{18432 m_{f_2}^4}  [ 424 m_\eta^{12} + 848 m_\eta^{10}(2 m_{f_2}^2-3m_G^2) + m_\eta^{8}(-57256 m_{f_2}^4 -12285m_{f_2}^2 m_G^2 +5712m_G^4)
 \\ \nonumber && + 4 m_\eta^6 (50896 m_{f_2}^6 +19441 m_{f_2}^4 m_G^2+1403 m_{f_2}^2 m_G^4- 1472m_G^6) -(m_{f_2}^2-m_G^2)^2
 \\ \nonumber && (46232 m_{f_2}^8-30339 m_{f_2}^6 m_G^2+19406 m_{f_2}^4 m_G^4-6363 m_{f_2}^2 m_G^6 +224 m_G^8) 
 \\ \nonumber && + m_\eta^4 (-290536 m_{f_2}^8+14306 m_{f_2}^6 m_G^2-171636 m_{f_2}^4 m_G^4+25658 m_{f_2}^2 m_G^6+2472 m_G^8) 
  \\ \nonumber && +4 m_\eta^2 (47080 m_{f_2}^{10}-50011 m_{f_2}^8 m_G^2-23767 m_{f_2}^6 m_G^4+33559 m_{f_2}^4 m_G^6-6873 m_{f_2}^2 m_G^8+ 12m_G^{10}) ]  \, .
\end{eqnarray}

\item The squared decay amplitudes of $| \mathrm{GG};2^{-+}\rangle \to \omega \omega$ can be expressed as
\begin{eqnarray}
 \mathcal{|M|}_{\omega \omega}^2 &\propto& \frac{m_G^4}{2034} (m_G^2-4 m_\omega^2)[-2304 f_\omega^2 {f^T_{\omega}}^2 m_\omega^2 m_G^2 (m_G^2+ m_\omega^2)
\\ \nonumber && +81{f^T_{\omega}}^4 m_G^4 (m_G^2+ m_\omega^2) +2048 f_\omega^4 (26m_\omega^6+ 11m_\omega^4 m_G^2)] \, .
\end{eqnarray}

\end{itemize}
\end{widetext}

\end{document}